\documentclass[aip,reprint]{revtex4-1}

\draft % marks overfull lines with a black rule on the right
\usepackage[pdftex]{graphicx}
\usepackage{textcomp}
\usepackage{hyperref}
\usepackage{amssymb}
\usepackage{amsmath}
\usepackage{ifthen} % provides \ifthenelse test  
\usepackage{xifthen} % provides \isempty test

\newcommand{\eq}[1]{equation~(\ref{#1})}
\newcommand{\fig}[2][]{%
\ifthenelse{\isempty{#1}}
{Fig.~\ref{#2}}% if no subfigure is given
{Fig.~\ref{#2}(#1)}% else
}

\begin{document}

%\preprint{}

\title{Damping of metallized bilayer nanomechanical resonators at room temperature} %Title of paper



\author{Maximilian J. Seitner}
\author{Katrin Gajo}
\author{Eva M. Weig}
\email[]{eva.weig@uni-konstanz.de}

%\thanks{}
\affiliation{University of Konstanz, Department of Physics, 78457 Konstanz, Germany}


\date{\today}

\begin{abstract}
We investigate the influence of gold thin-films subsequently deposited  on a set of initially bare, doubly clamped, high-stress silicon nitride string resonators at room temperature. Analytical expressions for resonance frequency, quality factor and damping for both in- and out-of-plane flexural modes of the bilayer system are derived, which allows for the determination of effective elastic parameters of the composite structure from our experimental data. We find the inverse quality factor to scale linearly with the gold film thickness, indicating that the overall damping is governed by losses in the metal. Correspondingly, the mechanical linewidth increases by more than one order of magnitude compared to the bare silicon nitride string resonator. Furthermore, we extract mechanical quality factors of the gold film for both flexural modes and show that they can be enhanced by complete deposition of the metal in a single step, suggesting that surface and interface losses play a vital role in metal thin-films.
\end{abstract}


\maketitle %\maketitle must follow title, authors, abstract and \pacs

Nanoelectromechanical systems (NEMS) enable both, the study of fundamental physical effects and future applications as integrated devices, e.g. in the field of ultra sensitive sensor technology\cite{2005_Ekinci_Rev}. As an important representative of such structures, freely suspended nanomechanical string resonators are exploited for their remarkable mechanical properties, providing high quality factors even at room temperature\cite{2006_Verbridge_JAP,2010_Quirin_PRL}. The overall performance of nanomechanical resonators is governed by their mechanical damping, originating from the combination of several fundamentally different loss mechanisms\cite{2014_Mohanty_physrep}. In general, extrinsic and intrinsic losses are distinguished, both of which have been studied extensively during the last decades. Intrinsic losses, such as thermoelasticity\cite{2000_Lifshitz_PRB}, clamping loss\cite{2008_Ignacio_PRB,2014_Rieger_NatComm}, surface\cite{2000_Yang_APL,2014_Villanueva_arxiv} and volume defects such as two-level system (TLS) damping\cite{2002_Pohl_Rev,2005_Vacher_PRB,2014_Faust_PRB} can be partially mitigated by intelligent resonator design. Extrinsic loss sources, like transduction mediated damping\cite{1999_Cleland_Sens,2012_Rieger_APL} can be reduced, e.g. through elimination of metallization layers on the resonant structure or the use of different detection techniques, such as optical schemes. However, hybrid nanostructures frequently rely on such metallization films, providing functionalization for coupling mechanical resonators to other degrees of freedom\cite{2008_Regal_NatPhys,2012_Kalman_NanoLett,2014_Kamra_PRB}. For those purposes it is inevitable to acquire a deeper understanding of the metal's impact on the overall damping of the system\cite{2005_Sandberg_JMM,2006_Beil_EL,2011_Sosale_JMM,2012_Yu_PRL} as well as its influence on elastic parameters\cite{2014_Hocke_arxiv}, especially at room temperature.\\
Here we present a thorough investigation of metallized bilayer nano-resonators based on doubly clamped, high-stress silicon nitride string resonators at room temperature. Starting from the basic assumptions of Euler-Bernoulli beam theory\cite{timoshenko}, we develop analytical expressions for the in- and out-of-plane harmonic eigenmodes of pre-stressed bilayer systems, which allows us to quantify their elastic constants. Furthermore, a model for the metal-induced quality factor is developed. Comparing the theoretical findings to our experimental data, the vast influence of the metal coating to the bare, non-metallized resonant structure becomes apparent. In the future, the resulting quantitative understanding will allow for the engineering of resonators with pre-defined frequency and quality factor by choosing the most suitable bilayer composition and resonator dimensions.

\begin{figure}[t]
\includegraphics[scale=0.23]{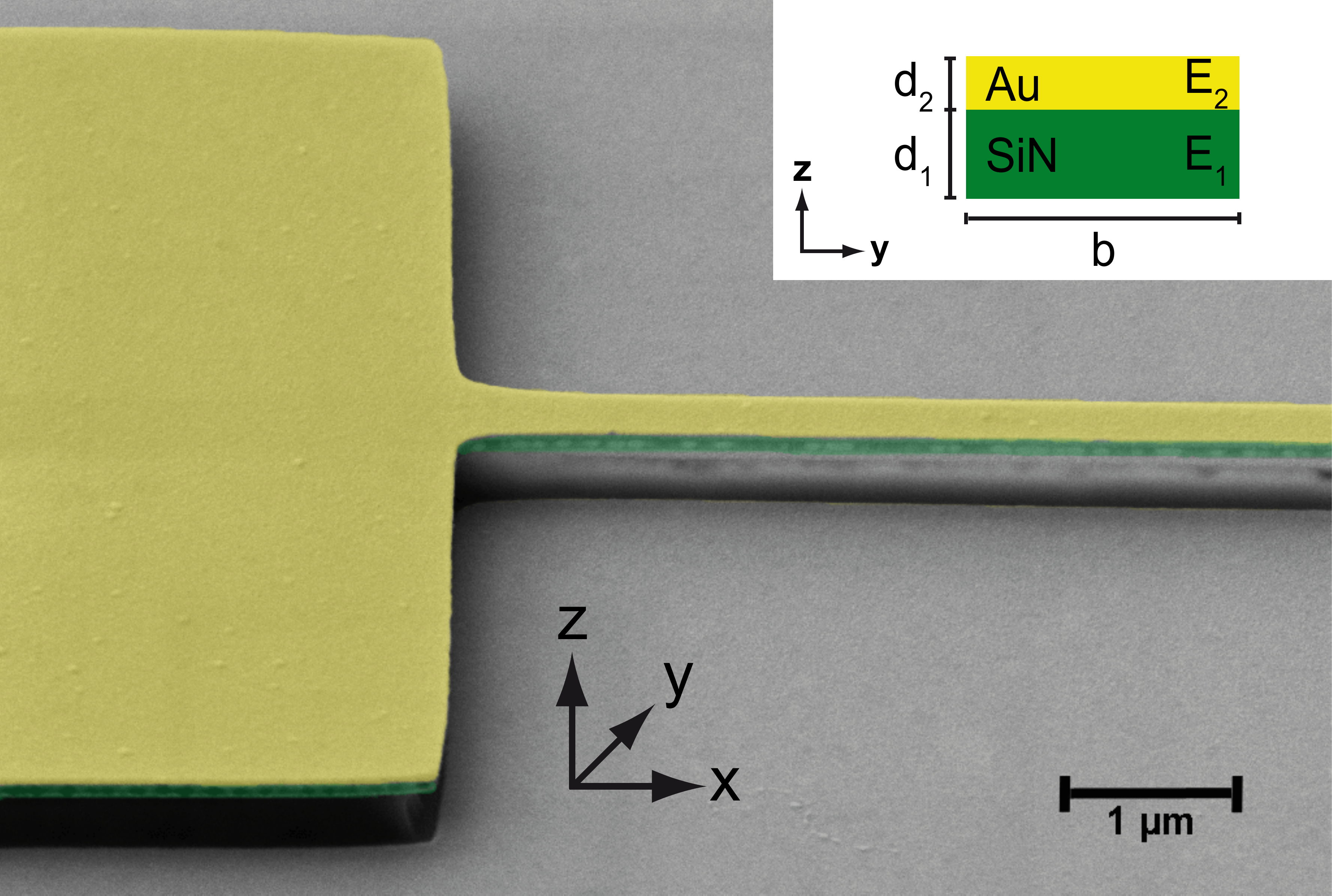}
\caption{\label{1}(color online). False color SEM micrograph of part of a doubly clamped silicon nitride string resonator covered with 50\,nm gold. Picture is taken under a tilted angle to point out the bilayer structure and the free suspension. Inset shows a schematic cross section along the resonator's width, introducing the geometric parameters width $b$=250\,nm, SiN thickness $d_1$=100\,nm and Au thickness $d_2$. $E_1$ and $E_2$ represent the respective Young's moduli.}
\end{figure}

\begin{figure}[htb]
\includegraphics[scale=0.3]{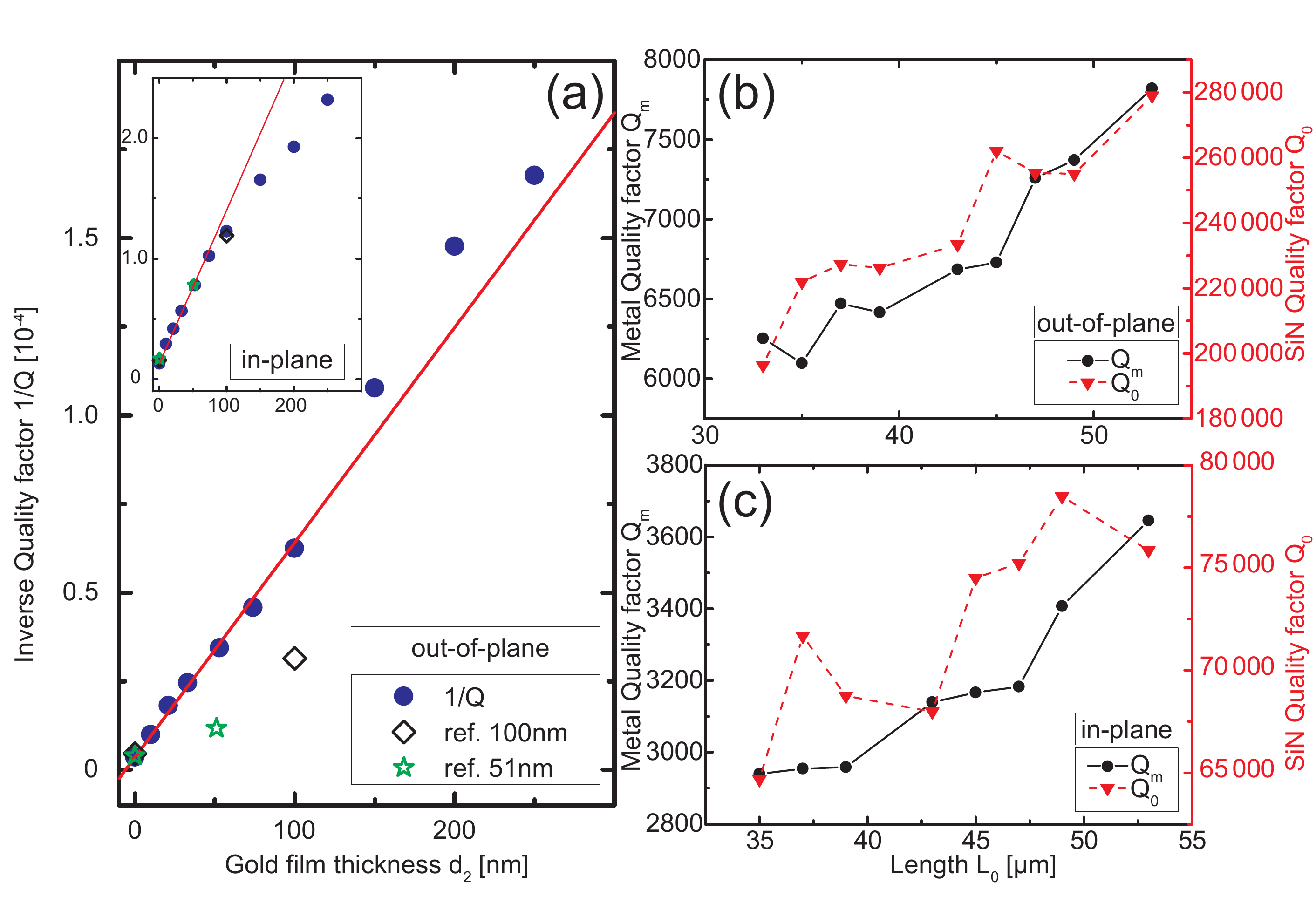}
\caption{\label{3} (color online). (a) Inverse Q-factor versus gold film thickness, including reference samples with 51\,nm (green stars) and 100\,nm (black diamonds) thick gold films for out-of-plane fundamental flexural mode of a 53\,\textmu m long resonator. Red solid line depicts fit to the 5 first data points with $\beta<0.25$. Inset shows corresponding in-plane mode. (b,c) Metal Q-factor (black circles, left scale) vs. resonator length. Red triangles (right scale) show Q-factors of non-metallized resonators for comparison, respectively.}
\end{figure}

\begin{figure}[top]
\includegraphics[scale=0.3]{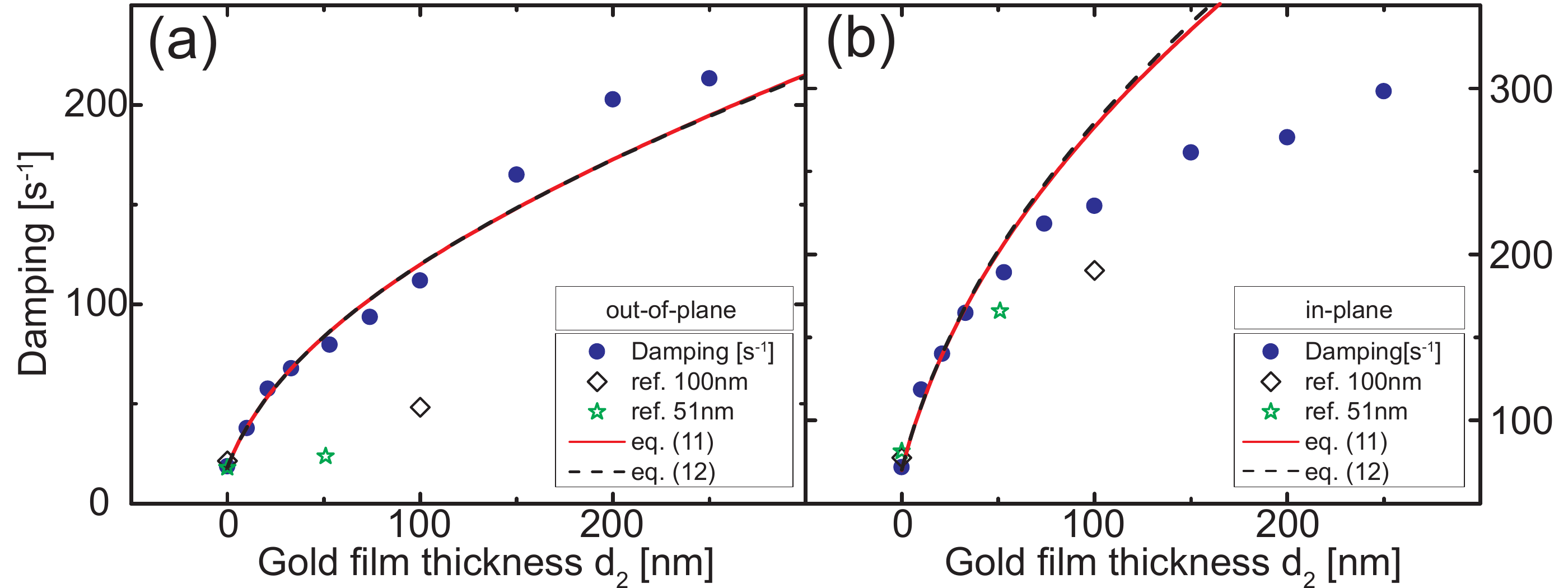}
\caption{\label{4}(color online). Damping, i.e. resonance linewidth, versus gold film thickness. (a) out-of-plane fundamental flexural mode of a 53\,\textmu m long resonator. (b) corresponding in-plane mode. Experimental data (blue dots) including reference samples with 51\,nm (green stars) and 100\,nm (black diamonds) thick gold films. Red solid (black dashed) lines have been calculated with no free parameters from eq.(11) and eq.(12), respectively.}
\end{figure}

A typical device is depicted in \fig{1}. We fabricate doubly clamped, freely suspended silicon nitride strings of different lengths on a silicon substrate in a top-down approach, using standard electron beam lithography and different dry and wet etching techniques. The LPCVD-grown amorphous silicon nitride on a sacrificial layer atop the silicon substrate is intrinsically pre-stressed, hence our resonators have a high inherent tensile stress of 830\,MPa\cite{2010_Quirin_PRL}. The bare silicon nitride strings have a thickness of $d_1$=100\,nm and a width of $b$=250\,nm, while their length varies from $L_0$=33\,\textmu m to $L_0$=53\,\textmu m. The resonances of their flexural modes are measured in an optical interferometer setup, where we refer to the oscillation in z-direction as out-of-plane (oop) and in y-direction as in-plane (ip) mode. After this first characterization, we successively deposit gold on the sample by electron beam evaporation to sequentially increase the total thickness $d_2$ of the metal. A 3\,nm adhesion layer of chromium is neglected in the following analysis\cite{2014_Mohanty_physrep}. For each gold film thickness we subsequently record the resonances of the now bilayer system and extract the characteristic oscillation parameters from Lorentzian fits, again using optical interferometric detection. Consequently we end up with results for nine different gold layer thicknesses on the same sample, as well as the data for the non-metallized SiN strings. Those results are evaluated according to a theoretical model of doubly clamped, pre-stressed bilayer nanomechanical resonators (see Supplemental Material\cite{SI_Max_APL} for detailed derivation).
Starting from the basic assumptions of Euler-Bernoulli beam theory\cite{timoshenko,Mech_Mat,2007_Wang_MMM}, we can express the bilayer system's resonant frequencies for both in- and out-of-plane flexural mode polarizations as\cite{SI_Max_APL} 

\begin{equation}
\label{f0}
f_{0,j}=\frac{j^2\pi}{2L^2}\sqrt{\frac{(EI)_{\rm eff}}{(\rho A)_{\rm eff}}}\sqrt{1+\frac{(\sigma A)_{\rm eff}L^2}{j^2(EI)_{\rm eff}\pi ^2}}
\end{equation}
with the harmonic mode index $j=1,2,3,...$ and effective material and elastic parameters $(\rho A)_{\rm eff}$, $(\sigma A)_{\rm eff}$ and $(EI)_{\rm eff}$. The effective density is
\begin{equation}
\label{rho}
(\rho A)_{\rm eff}=\frac{\rho _1d_1+\rho _2d_2}{d_1+d_2}A=\widetilde{\rho}A
\end{equation}
and the effective tensile stress reads
\begin{equation}
\label{sigma}
(\sigma A)_{\rm eff}=\frac{\sigma _1d_1+\sigma _2d_2}{d_1+d_2}A=\widetilde{\sigma}A
\end{equation}
where $A=b(d_1+d_2)$ equals the cross section area of the bilayer (see inset of \fig{1}). The effective bending rigidity depends on the polarization of the flexural mode\cite{SI_Max_APL}, namely

\begin{multline}
\label{oop}
(EI)_{\rm eff,oop}=\\
b\frac{E_1^2d_1^4+2E_1E_2d_2(2d_1^3+2d_1d_2^2+3d_1^2d_2)+E_2^2d_2^4}{12(E_1d_1+E_2d_2)}\\
=E_1I_{\rm eff}^{\rm oop}
\end{multline}
for out-of-plane and
\begin{equation}
\label{ip}
(EI)_{\rm eff,ip}=\frac{b^3(E_1d_1+E_2d_2)}{12}=E_1I_{\rm eff}^{\rm ip}
\end{equation}
for in-plane oscillations. For low harmonic modes of high aspect ratio beams exhibiting high tensile stress
\begin{equation}
\label{simp}
\frac{(\sigma A)_{\rm eff}L^2}{j^2(EI)_{\rm eff}\pi ^2}\gg 1 \,\,\,\,\, ,
\end{equation}
and \eq{f0} reduces to
\begin{equation}
\label{string}
f_{0,j}\approx\frac{j}{2L}\sqrt{\frac{\widetilde{\sigma}}{\widetilde{\rho}}}=\frac{j}{2L}\sqrt{\frac{\sigma _1d_1+\sigma _2d_2}{\rho _1d_1+\rho _2d_2}} \,\,\,\,\, .
\end{equation}
\begin{figure}[t]
\includegraphics[scale=0.37]{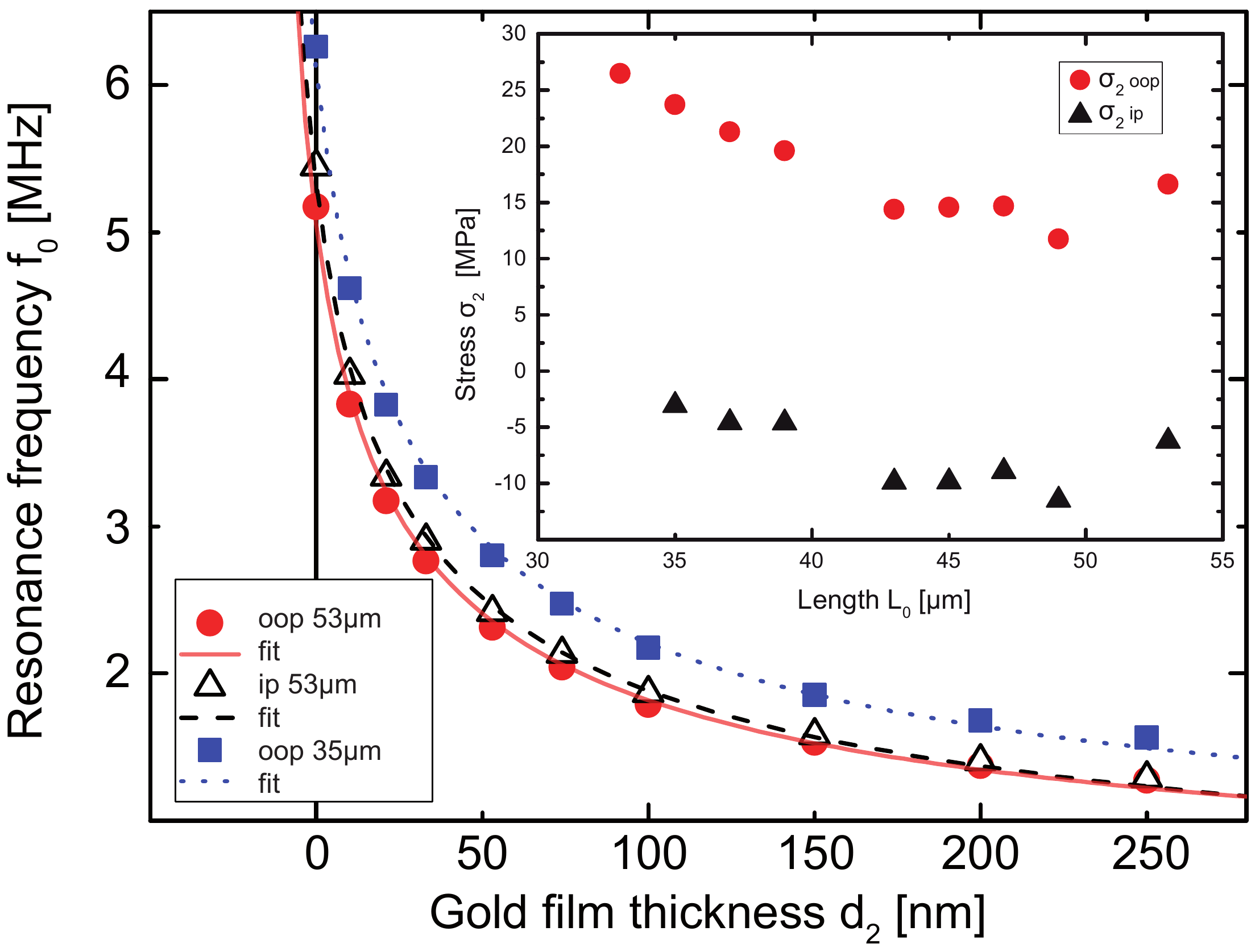}
\caption{\label{2}(color online). Decrease of resonance frequency as a function of gold film thickness, for the fundamental out-of-plane flexural modes of a 53\,\textmu m (red dots) and a 35\,\textmu m (blue squares) as well as the 53\,\textmu m long resonator's corresponding fundamental in-plane mode (open black triangles). Red solid, blue dotted and black dashed lines represent the fit of \eq{f0} to the data, respectively. The inset shows fitted tensile stress of gold films versus resonator length for in-plane (black triangles) and out-of-plane mode (red dots).}
\end{figure}
Using the established values for the material constants of our silicon nitride strings\cite{2010_Quirin_PRL}, density $\rho _1$=2.8\,g/$\rm cm^3$, Young's modulus $E_1$=160\,GPa, tensile stress $\sigma _1$=830\,MPa, and literature values for the gold films\cite{Springer_Materials},  $\rho _2$=19.3\,g/$\rm cm^3$ and $E_2$=78\,GPa, we are now able to fit the theoretical predictions to our experimental data. We used two fit parameters, the tensile stress $\sigma _2$ of the gold film, as well as the length of the resonator $L$. Note that $L$ does not coincide with the nominal length of the resonator $L_0$ as a result of the undercut of the clamping points during the HF wet etch\cite{2014_Rieger_NatComm} (c.f. \fig{1}). Exemplary results are depicted in \fig{2} showing the fundamental out-of-plane eigenfrequencies of a nominally $L_0$=53\,\textmu m and of a $L_0$=35\,\textmu m long resonator as well as the corresponding in-plane flexural mode of the 53\,\textmu m string as a function of gold film thickness. We find the data in good agreement with the theoretical predictions. The obtained values of $L$ consistently exceed $L_0$ with the exception of one data point, which is considered as outlier\cite{SI_Max_APL}. The effective elongation $(L-L_0)$ is more pronounced for shorter strings and for out-of-plane modes\cite{SI_Max_APL}, reflecting the larger impact of the underetched clamping region on vibrations perpendicular to the sample plane. The inset of \fig{2} displays the tensile stress in the gold film for resonators of different lengths extracted from the fit. We find positive values for $\sigma _2$ in the out-of-plane, and slightly negative values for the in-plane mode, both with a tendency towards smaller $\sigma _2$ for longer resonators. The overall magnitude of $\sigma _2$ lies well within the range of stress expected for e-beam evaporated gold films\cite{1960_Neugebauer_JAP}. This stress originates from the evaporation of the gold and should not depend on resonator length. The dependence of $\sigma _2$ on length and mode polarization remains topic of further investigation. However, since the extracted stress in the gold film is very small compared to the residual stress in the silicon nitride ($<4\,\%$) this effect may well be interpreted as an artifact in our fitting routine of \eq{f0} to \eq{ip}.\\
Metallizing nanomechanical resonators not only affects their resonant frequencies. The metal's deleterious impact is most pronounced in the otherwise high room temperature mechanical quality factor (Q-factor) of our silicon nitride string resonators. The past decades showed extensive studies on metal thin-film damping, mostly at low temperatures\cite{2014_Mohanty_physrep,2006_Beil_EL,2011_Sosale_JMM,2012_Yu_PRL,1999_Liu_PRB,2004_Hutchinson_APL,2005_Sandberg_JMM,2006_Feng_arxiv}. According to this framework, we describe the metal's impact on the inverse quality factor 1/Q. Here one should point out deliberately that by 1/Q we refer to the inverse quality factor and not the damping. We find 1/Q of a bilayer structure to equal\cite{1995_White_PRL,2014_Mohanty_physrep}
\begin{equation}
\label{iq}
\frac{1}{Q}=\frac{1}{1+\beta}\left(\frac{1}{Q_0}+\beta \frac{1}{Q_m}\right)
\end{equation}
where $Q_0$ represents the Q-factor of the non-metallized resonator and $Q_m$ an effective mechanical Q-factor of the metal film. The dimensionless factor $\beta =(E_2d_2)/(E_1d_1)$ depends on the ratio of the material's Young's moduli and thicknesses and can be interpreted as the ratio of the relative bilayer thicknesses within the framework of transformed sections\cite{Mech_Mat}. Since in our case $\beta$ is typically small, we can approximate \eq{iq} in a Taylor series expansion to first order in $\beta$
\begin{equation}
\label{q}
\frac{1}{Q}=\frac{1}{Q_0}+\frac{E_2}{E_1Q_0}\left(\frac{Q_0}{Q_m}-1\right)\frac{d_2}{d_1} \,\,\,\,\, .
\end{equation}
This equation clearly shows that the inverse quality factor of the bilayer system scales linearly with the gold film thickness $d_2$ in the limit of thin metal films, i.e. small $\beta$. This behavior is experimentally validated in \fig{3}. 
Here we plot the inverse mechanical Q-factor as a function of the gold film thickness $d_2$. Since \eq{q} holds only for small values of $\beta$, we consider only the first five data points for fitting (red solid line), such that $d_2\leq 50$\,nm which yields $\beta<0.25$. For thicker films our approximation breaks down and $1/Q$ begins to deviate from the first order expansion. Note that this effect gets more pronounced for higher harmonic modes\cite{SI_Max_APL}. From the fit of \eq{q} (red solid line) we extract values of $Q_m$, which are interpreted as effective Q-factors of the metal\cite{1999_Liu_PRB,2000_Craighead_JVS,2004_Hutchinson_APL,2006_Feng_arxiv}. As displayed in \fig{3}(b,c), $Q_m$ tends to increase approximately linear with increasing resonator length and therefore decreasing resonant frequency. This behavior is in qualitative agreement with that of the bare SiN strings (red triangles in \fig{3}(b,c)) and is already known from defect damping dominated systems\cite{2006_Verbridge_JAP,2010_Hoehne_PRB,2010_Venkatesan_PRB}, indicating that the bilayer system's damping is governed by defects in the metal. The contribution from other fundamental loss mechanisms, such as thermoelasticity\cite{2000_Lifshitz_PRB}, is significantly smaller than the measured $Q_m$\cite{2007_Prabhakar_JMM}. Overall, the observed effective room temperature mechanical quality factors of the gold films of several thousands agree well with literature findings from different metallic systems\cite{1999_Liu_PRB,2000_Craighead_JVS,2004_Hutchinson_APL,2006_Feng_arxiv}.\newline
In order to investigate the metal induced $Q_m$ in a more detailed way, we examine two reference samples where gold films of 51\,nm and 100\,nm thickness (green stars and black diamonds in \fig{3}(a)) have been evaporated on the sample in a single step, respectively. These structures yield metal quality factors up to $Q_m$=20\,000 for out-of-plane and $Q_m$=4\,000 for in-plane flexural fundamental modes. Compared to the case of subsequently evaporated gold films single step evaporation results in a significant improvement of the Q-factor of the out-of-plane oscillation direction, whereas the Q-factor of the in-plane oscillation remains rather unaffected. We attribute this to the impact of surface losses\cite{2014_Villanueva_arxiv}. Subsequent evaporation of several gold layers creates additional interfaces. Furthermore, the sequential investigation implies exposing the device to air, which leads to additional contamination layers along the out-of-plane oscillation direction, yielding more damping and therefore lower Q-factors for out-of-plane vibrations. Contrary, the additional interfaces comprehensively have less impact on in-plane vibrations.\newline
Since the metal's quality factors $Q_m$ are significantly smaller than the bare silicon nitride $Q_0$'s, one can approximate the second term of \eq{q} even further
\begin{equation}
\label{apq}
\frac{E_2}{E_1Q_0}\left(\frac{Q_0}{Q_m}-1\right)\frac{d_2}{d_1}\approx \frac{1}{Q_m}\frac{E_2}{E_1}\frac{d_2}{d_1}
\end{equation}
The damping constant, defined in units of frequency, corresponds to the linewidth of the mechanical resonance and can hence be expressed as
\begin{equation}
\label{gamma}
\gamma_j=\frac{f_{0,j}}{Q}=f_{0,j}\left(\frac{1}{Q_0}+\frac{1}{Q_m}\frac{E_2}{E_1}\frac{d_2}{d_1}\right) \,\,\,\,\, .
\end{equation}
Using the experimentally determined and material parameters from above, the bilayer system's damping can now be calculated from \eq{gamma} with no free parameters. \fig{4} displays the resulting damping curve (red line), along with the experimental data. We find excellent agreement between experiment and theoretical assumption for the first five data points of flexural modes where the first order approximation of $\beta$ still holds. For the case of the strongly stressed string, \eq{gamma} can be approximated using \eq{simp}, yielding
\begin{equation}
\label{stringgamma}
\gamma_j=\frac{j}{2L}\sqrt{\frac{\sigma _1d_1+\sigma _2d_2}{\rho _1d_1+\rho _2d_2}}\left(\frac{1}{Q_0}+\frac{1}{Q_m}\frac{E_2}{E_1}\frac{d_2}{d_1}\right) \,\,\,\,\, .
\end{equation}
 The dashed line in \fig{4} has been obtained from the high-stress approximation in \eq{stringgamma}. It coincides with the red curve, confirming that the elastic properties of the metallized bilayer system are indeed dominated by the high effective tensile stress in the material. Taking into account that the total damping amounts to the sum of all contributing damping mechanisms\cite{2014_Mohanty_physrep}, \eq{gamma} and \eq{stringgamma} can be interpreted as the sum of the damping in SiN and gold, respectively. Indeed, the first term in \eq{stringgamma} accounts for the damping of the non-metallized SiN string resonator. The second term is governed by the quality factor assigned to the metal film as well as the ratio of the effective thicknesses\cite{Mech_Mat}.
\linebreak
In conclusion, we have investigated the mechanical properties of gold-coated silicon nitride string resonators at room temperature. Using analytical expressions for resonant frequency, quality factor and damping of bilayer structures we have been able to show that the damping of metallized bilayer nanomechanical string resonators is governed by defect losses in the metal film and scales with its thickness. Comparison of the subsequently evaporated gold coatings with reference samples evaporated in a single step revealed that surface losses play an important role in metal dissipation. The presented thorough analysis enables engineering metallized bilayer resonators with custom eigenfrequency and quality factor by adjusting the resonator dimensions and metal thickness. This is of utmost importance for future applications of metallized resonators in hybrid nanomechanical systems where a good understanding of the mechanical system is required, for example the functionalization for biosensing applications\cite{2011_Arlett_Natnano} or the electromagnetic coupling of NEMS to cold atoms\cite{2012_Kalman_NanoLett,2014_Darazs_PRL}. Furthermore, the above can easily be extended to other metals, enabling to further reduce the metal's damping contribution\cite{2010_Collin_JAP} or to open up routes towards coupling to other degrees of freedom for hybrid nanosystems by using superconducting\cite{2008_Regal_NatPhys} or magnetic\cite{2014_Kamra_PRB} metallizations.

\begin{acknowledgments}
Financial support by the Deutsche Forschungsgemeinschaft via the collaborative research center SFB\,767 is gratefully acknowledged. We thank James A. Dorman for critically reading the manuscript.
\end{acknowledgments}

% If in two-column mode, this environment will change to single-column format so that long equations can be displayed. 
%\begin{widetext}
%$$\mbox{put long equation here}$$
%\end{widetext}

% Figures should be put into the text as floats. 
% Use the graphics or graphicx packages (distributed with LaTeX2e).
% Put the label that you will use with \ref{} command in the braces of the \label{} command.
%
% \begin{figure}
% \end{figure}

% \begin{table}
% \end{tabular}
% \end{table}

%\begin{acknowledgments}
% Put your acknowledgments here.
%\end{acknowledgments}

%\bibliography{metal_paper}

%\bibliography{metal_paper}

%

\renewcommand{\thefigure}{S\arabic{figure}}
 \renewcommand{\theequation}{S\arabic{equation}}
 \renewcommand{\thetable}{S\arabic{table}}
 
 \renewcommand{\eq}[1]{equation~(\ref{#1})}
 \renewcommand{\citenumfont}[1]{S#1}
 \renewcommand{\bibnumfmt}[1]{[S#1]}
 
 \setcounter{figure}{0}
 \setcounter{equation}{0}
 
 \onecolumngrid
 \clearpage

 \renewcommand{\refname}{Supplemental Material References}

\section*{Supplemental Material to "Damping of metallized bilayer nanomechanical resonators at room temperature"}

\pagestyle{empty}

\vspace{1cm}

\section{Theoretical derivation of flexural vibrations of a bilayer beam subjected to residual stresses}
The theoretical modelling is performed using the framework of Euler-Bernoulli elastic beam theory\cite{timoshenko2}.
\subsection{Geometry}

We consider a prismatic, rectangular double layer beam of Length $L$, where the indices 1,\,2 refer to layer\,1 and layer\,2, respectively. The geometry is defined equivalently to Fig.\,1 of the main article and depicted in Fig.\,S1.

\begin{figure}[htb]
\label{s1}
\includegraphics{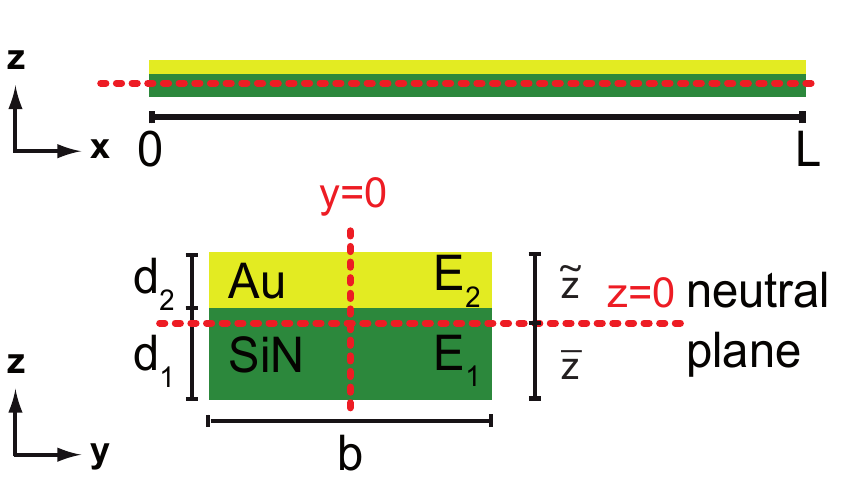}
\caption{Geometry of the bilayer structure following the coordinate system of Fig.\,1 in the main article. Dashed red lines illustrate the neutral planes of in- and out-of-plane vibrations which are not necessarily in the mid-plane.}
\end{figure}

We define the x-axis to coincide with the neutral plane of the out-of-plane vibration such that
\begin{gather}
\label{l1}
{\rm layer\,1}:\,\, -\bar z < z < d_1-\bar z \\
\label{l2}
{\rm layer\,2}:\,\,  d_1-\bar z < z < d_1+d_2-\bar z \\
-\frac{b}{2}<y<\frac{b}{2}
\end{gather}

\subsection{Assumptions}
The following basic assumptions hold for the complete derivation including both, the out-of-plane and the in-plane vibration case.

\begin{itemize}
\item Each layer is homogenous and isotropic and obeys Hooke's law (elastic deformation regime)
\item Transverse planes remain transverse (Euler-Bernoulli hypothesis):\\ This implies the existence of a neutral plane
\item Consider pure bending modes\\ (neglect shear deformation and rotary inertia)
\item Both layers inhere residual stress:\\ Stresses are assumed to be completely released along the y- and z-direction;\\ uniformly distributed along the x-direction in each layer, respectively
\item Double layer system:\\ Assume perfect adhesion of the laminated layers across the entire interface
\end{itemize}

\subsection{Out-of-plane vibration}

\subsubsection{Neutral plane}

The neutral plane is defined by the condition that the axial force from pure bending acting on the entire cross-section has to vanish
\begin{equation}
\label{s0}
\int_{area} \sigma _{\rm bend} \, dA \equiv 0
\end{equation}
Using Hooke's law $\sigma_{\rm bend}=E_i \, \epsilon _{\rm bend}$ with Youngs moduli $E_i$\,(i=1,2) and the strain-displacement relation $\epsilon _{\rm bend}=-z/\kappa$ with bending curvature $g=1/\kappa = (\partial ^2w)/(\partial x^2)$, where $w=w(z)$ represents a displacement in z-direction, we get
\begin{equation}
-\int_{area\,A_1}E_1\frac{z}{\kappa}\,dA_1-\int_{area\,A_2}E_2\frac{z}{\kappa}\,dA_2 \equiv 0
\end{equation}
In order to integrate this equation we choose a coordinate system in which the x-axis coincides with the neutral axis (compare \eq{l1} and \eq{l2}), i.e. we use the transformation $\widetilde z=z-\bar z$.
\begin{gather*}
-\frac{E_1}{\kappa}\int_{-b/2}^{b/2}dy\int_0^{d_1}(z-{\bar z}) \,d\widetilde z -\frac{E_2}{\kappa}\int_{-b/2}^{b/2}dy\int_{d_1}^{d_1+d_2}(z-{\bar z})\,d\widetilde z=\\
=-E_1\frac{b}{\kappa}\left(\left[\frac{1}{2}z^2\right]_{0}^{d_1}-\bar z d_1\right)-E_2\frac{b}{\kappa}\left(\left[\frac{1}{2}z^2\right]_{d_1}^{d_1+d_2}-\bar z d_2\right) \equiv 0
\end{gather*}
\begin{equation}
\label{np}
\implies \bar z=\frac{E_1d_1^2+E_2d_2^2+2E_2d_1d_2}{2(E_1d_1+E_2d_2)}
\end{equation}

\subsubsection{Displacement}

We now consider a displacement of an infinitesimal volume element in x- and z-direction (see Fig.\,S2). The displacement of the neutral axis along the x- and z-axis is then given by
\begin{gather*}
u_x(x,y,z)=u_x(x,z)=-z \sin \Theta\\
u_z(x,y,z)=u_z(x,z)=w(x,z)=w(x)
\end{gather*}
For small displacements, i.e. small angles $\Theta$ one can approximate $\sin \Theta \approx \Theta \approx \tan \Theta$ and therefore
\begin{equation}
\label{ux}
u_x(x,z)=-z\frac{dw(x)}{dx}
\end{equation}

\begin{figure}[htb]
\label{s2}
\includegraphics{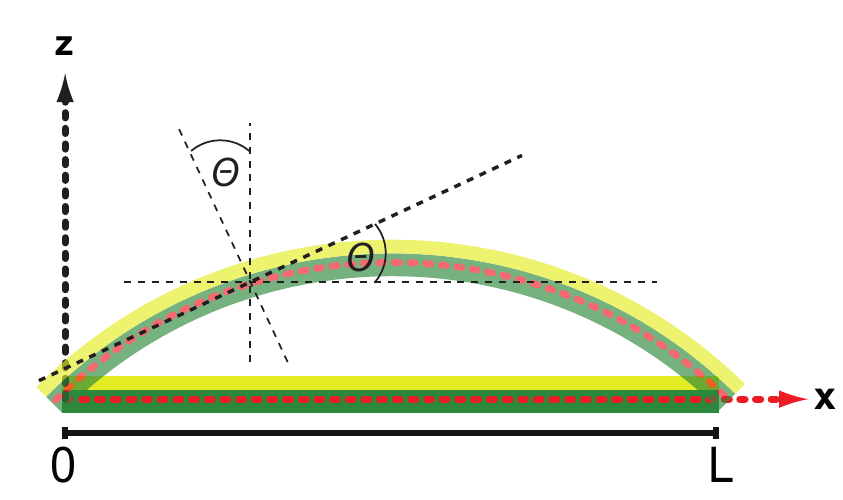}
\caption{Illustration of the displacement of the bilayer beam along z-direction. The neutral plane of the displaced beam defines the x-axis.}
\end{figure}
For the sake of completeness it should be noted that in addition in laminated beam theory\cite{2007_Wang_MMM2}, an axial displacement of the material axis $u_{x0}=u(x)$ is frequently considered as depicted in Fig.\,S3.

\begin{figure}[htb]
\label{s3}
\includegraphics[scale=1.5]{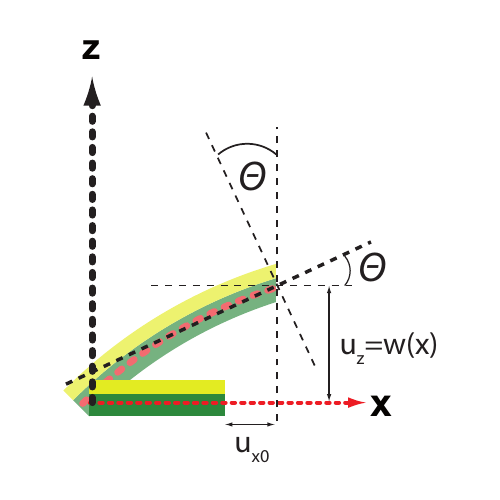}
\caption{Axial displacement $u_{x0}$ in laminated beam theory.}
\end{figure}

\begin{gather}
u_x(x,y,z)=u_{x0}(x)+u_{x}(x,z)=u(x)-z\frac{dw(x)}{dx} \tag{S7.a} \label{uxa}\\
u_z(x,y,z)=u_z(x,z)=w(x)
\end{gather}
So for the bilayer system, this implies implicitly:
\begin{gather}
\label{disp}
u_{1x}(x,y,z)=u(x)-z\frac{dw(x)}{dx} \,\,\,\,\, {\rm for} -\bar z < z < d_1-\bar z\\
u_{2x}(x,y,z)=u(x)-z\frac{dw(x)}{dx} \,\,\,\,\, {\rm for} \,\,\,d_1-\bar z < z < d_1+d_2-\bar z\\
u_z(x,y,z)=w(x) \,\,\,\,\, {\rm for} -\bar z < z < d_1+d_2-\bar z
\end{gather}
where $u_{1x}$ and $u_{2x}$ refer to the displacement of layer\,1 and layer\,2 along the x-axis, respectively, and $u_{1z}=u_{2z}\equiv u_{z}$.

\subsubsection{Strain}
The Green-St.\,Venant strain tensor $\epsilon_{ij}$ is defined as
\begin{equation}
\epsilon_{ij}=\frac{1}{2}\left(\frac{\partial u_i}{\partial r_j}+\frac{\partial u_j}{\partial r_i}+\sum\limits_{k=1}^{3}\frac{\partial u_k}{\partial r_i}\frac{\partial u_k}{\partial r_j}\right)
\end{equation}
For the case of uniaxial strain along the x-axis this corresponds to
\begin{equation}
\epsilon_{x}:=\epsilon_{xx}=\frac{\partial u_x}{\partial x}+\frac{1}{2}\left(\left(\frac{\partial u_x}{\partial x}\right)^2+\left(\frac{\partial u_z}{\partial z}\right)^2\right)
\end{equation}
In standard Euler-Bernoulli theory, only small deformations are considered and the strain is conveniently linearized
\begin{equation}
\label{epsilon}
\epsilon_{x}=\frac{\partial u_x}{\partial x}
\end{equation}
In the so-called moderately nonlinear regime the slopes $(\partial u_x)/(\partial x)$ and $(\partial u_z)/(\partial z)$ are still small so that the nonlinearity reduces in lowest order to
\begin{equation}
\label{epsilonx}
\epsilon_{x}=\frac{\partial u_x}{\partial x}+\frac{1}{2}\left(\frac{\partial u_z}{\partial z}\right)^2 \tag{S14.a}
\end{equation}
Using the displacement \eq{uxa} we obtain for the simple beam:
\begin{equation}
\label{ex}
\epsilon_{x}=\frac{\partial u_x}{\partial x}+\frac{1}{2}\left(\frac{\partial u_z}{\partial z}\right)^2\\
=\frac{\partial u(x)}{\partial x}-z\frac{\partial ^2w(x)}{\partial x^2}+\frac{1}{2}\left(\frac{\partial w(x)}{\partial x}\right)^2
\end{equation}
The first term of \eq{ex} describes local elongation (from axial strain), the second one local bending of the beam (from bending strain) and the last one additional local elongation of the beam caused by bending (second order effect and nominally small).\newline
For the bilayer system this corresponds to
\begin{gather}
\epsilon_{1x}=\frac{\partial u(x)}{\partial x}-z\frac{\partial ^2w(x)}{\partial x^2}+\frac{1}{2}\left(\frac{\partial w(x)}{\partial x}\right)^2 \,\,\,\,\, {\rm for} -\bar z < z < d_1-\bar z\\
\epsilon_{2x}=\frac{\partial u(x)}{\partial x}-z\frac{\partial ^2w(x)}{\partial x^2}+\frac{1}{2}\left(\frac{\partial w(x)}{\partial x}\right)^2 \,\,\,\,\, {\rm for} \,\,\,d_1-\bar z < z < d_1+d_2-\bar z
\end{gather}
The stress-strain relation is given by Hooke's law for the two materials:
\begin{gather}
\sigma_{1x}=E_1\left[\frac{\partial u(x)}{\partial x}-z\frac{\partial ^2w(x)}{\partial x^2}+\frac{1}{2}\left(\frac{\partial w(x)}{\partial x}\right)^2\right]+\sigma _1 \,\,\,\,\, {\rm for} -\bar z < z < d_1-\bar z\\
\sigma_{2x}=E_2\left[\frac{\partial u(x)}{\partial x}-z\frac{\partial ^2w(x)}{\partial x^2}+\frac{1}{2}\left(\frac{\partial w(x)}{\partial x}\right)^2\right]+\sigma _2 \,\,\,\,\, {\rm for} \,\,\,d_1-\bar z < z < d_1+d_2-\bar z
\end{gather}
Here we introduced the individual residual pre-stress $\sigma _{1/2}$ in both layers, respectively. Note that the bending-induced stress $\sigma_{\rm ni}$\,(n=1,2\,;\,i=xx,yy,zz) needs to be distinguished from the residual pre-stress $\sigma_{\rm n}$\,(n=1,2).

\subsubsection{Forces and moments}

In this section we use the previous findings and definitions to calculate the axial force and the bending moment of the bilayer structure.
We define the axial force along the x-direction as follows:
\begin{gather*}
N_x=\int_{area\,A_1}\sigma _{1x}\,dA+\int_{area\,A_2}\sigma _{2x}\,dA\\
=\int_{-b/2}^{b/2}dy\int_{-\bar z}^{d_1-\bar z}\sigma _{1x}\,dz+\int_{-b/2}^{b/2}dy\int_{d_1-\bar z}^{d_1+d_2-\bar z}\sigma _{2x}\,dz\\
=b\int_{-\bar z}^{d_1-\bar z}\left[E_1\left(\frac{\partial u(x)}{\partial x}-z\frac{\partial ^2w(x)}{\partial x^2}+\frac{1}{2}\left(\frac{\partial w(x)}{\partial x}\right)^2\right)+\sigma _1\right]\,dz\\
+b\int_{d_1-\bar z}^{d_1+d_2-\bar z}\left[E_2\left(\frac{\partial u(x)}{\partial x}-z\frac{\partial ^2w(x)}{\partial x^2}+\frac{1}{2}\left(\frac{\partial w(x)}{\partial x}\right)^2\right)+\sigma _2\right]\,dz\\
=b\{\left(\frac{\partial u(x)}{\partial x}+\frac{1}{2}\left(\frac{\partial w(x)}{\partial x}\right)^2\right)\left[E_1\left((d_1-\bar z)+\bar z\right)+E_2\left((d_1+d_2-\bar z)-(d_1-\bar z)\right)\right]\\
+\left[\sigma _1\left((d_1-\bar z)+\bar z\right)+\sigma _2\left((d_1+d_2-\bar z)-(d_1-\bar z)\right)\right]\\
-\frac{1}{2}\frac{\partial ^2w(x)}{\partial x^2}[E_1((d_1-\bar z)^2-(-\bar z)^2+E_2((d_1+d_2-\bar z)^2-(d_1-\bar z)^2)]\}
\end{gather*}
Plugging in the definition of the neutral plane $\bar z$ (\eq{np}) and doing more straight forward simplifications we end up with
\begin{equation}
\label{nx}
N_x=b\left\{\left(\frac{\partial u(x)}{\partial x}+\frac{1}{2}\left(\frac{\partial w(x)}{\partial x}\right)^2\right)\left[E_1d_1+E_2d_2\right]+\left[\sigma _1d_1+\sigma _2d_2\right]\right\}
\end{equation}
The bending moment in y-direction reads:
\begin{gather*}
M_y=\int_{area\,A_1}(\vec{r}\times \vec{\sigma})_y\,dA+\int_{area\,A_2}(\vec{r}\times \vec{\sigma})_y\,dA\\
=\int_{-b/2}^{b/2}dy\int_{-\bar z}^{d_1-\bar z}z\sigma _{1x}\,dz+\int_{-b/2}^{b/2}dy\int_{d_1-\bar z}^{d_1+d_2-\bar z}z\sigma _{2x}\,dz\\
=b\int_{-\bar z}^{d_1-\bar z}z\left[E_1\left(\frac{\partial u(x)}{\partial x}-z\frac{\partial ^2w(x)}{\partial x^2}+\frac{1}{2}\left(\frac{\partial w(x)}{\partial x}\right)^2\right)+\sigma _1\right]\,dz\\
+b\int_{d_1-\bar z}^{d_1+d_2-\bar z}z\left[E_2\left(\frac{\partial u(x)}{\partial x}-z\frac{\partial ^2w(x)}{\partial x^2}+\frac{1}{2}\left(\frac{\partial w(x)}{\partial x}\right)^2\right)+\sigma _2\right]\,dz\\
=b\{\frac{1}{2}\left(\frac{\partial u(x)}{\partial x}+\frac{1}{2}\left(\frac{\partial w(x)}{\partial x}\right)^2\right)\left[E_1((d_1-\bar z)^2-(-\bar z)^2)+E_2((d_1+d_2-\bar z)^2-(d_1-\bar z)^2)\right]\\
+\frac{1}{2}\left[\sigma _1((d_1-\bar z^2)-(-\bar z)^2)+\sigma _2((d_1+d_2-\bar z)^2-(d_1-\bar z)^2)\right]\\
-\frac{1}{3}\frac{\partial ^2w(x)}{\partial x^2}\left[E_1((d_1-\bar z)^3-(-\bar z)^3)+E_2((d_1+d_2-\bar z)^3-(d_1-\bar z)^3)\right]
\}
\end{gather*}
Following the appendix of Wang et al.\cite{2007_Wang_MMM2} the result can be expressed by parametrizing the layer coordinates as
\begin{gather}
\label{newy}
z_0=-\bar z\\
z_1=d_1-\bar z\\
z_2=d_1+d_2-\bar z
\end{gather}
Therefore the bending moment can be rewritten
\begin{equation}
\label{mz}
M_y=b\left\{\widetilde B\left[\frac{\partial u(x)}{\partial x}+\frac{1}{2}\left(\frac{\partial w(x)}{\partial x}\right)^2\right]-\widetilde D\frac{\partial ^2w(x)}{\partial x^2}+\widetilde M\right\}
\end{equation}
as well as the axial force $N_x$(\eq{nx})
\begin{equation}
\label{newnx}
N_x=b\left\{\widetilde A\left[\frac{\partial u(x)}{\partial x}+\frac{1}{2}\left(\frac{\partial w(x)}{\partial x}\right)^2\right]+\widetilde N\right\}
\end{equation}
with the tension stiffness
\begin{equation}
\widetilde A= E_1d_1+E_2d_2
\end{equation}
and the tension-bending coupling stiffness
\begin{gather}
\widetilde B= \frac{1}{2}E_1(z_1^2-z_0^2)+\frac{1}{2}E_2(z_2^2-z_1^2)=0\\
\widetilde D= \frac{1}{3}E_1(z_1^3-z_0^3)+\frac{1}{3}E_2(z_2^3-z_1^3)
\end{gather}
where $\widetilde B$ is equal to zero due to the choice of the coordinate system that the x-axis lies in the neutral plane (compare \eq{l1} and \eq{l2}).
The residual force reads
\begin{equation}
\widetilde N= \sigma _1d_1+\sigma _2d_2
\end{equation}
and the residual moment
\begin{equation}
\widetilde M= \frac{1}{2}\sigma _1(z_1^2-z_0^2)+\frac{1}{2}\sigma _2(z_2^2-z_1^2)
\end{equation}

\subsubsection{Equilibrium equations}

The equilibrium equations of the bilayer system can be derived in complete analogy to the well known single layer Euler-Bernoulli beam\cite{timoshenko2} by considering a small volume element $dx\,dA$ of Fig.\,S1. and Fig.\,S2. in the x-z-plane.\newline
The equilibrium condition for axial forces along the x-direction reads
\begin{gather*}
N_x(x+dx)-N_x(x)=0\\
N_x(x)+dx\frac{\partial N_x(x)}{\partial x}-N_x(x)=0\\
\iff \frac{\partial N_x(x)}{\partial x} \equiv 0
\end{gather*}
Along the z-direction the shear force $V_z$ and the axial force $N_z$ compensate each other to zero
\begin{equation*}
[V_z(x+dx)-V_z(x)]+[N_z(x+dx)-N_z(x)]=0
\end{equation*}
where
\begin{equation*}
[V_z(x+dx)-V_z(x)]=V_z(x)+dx\frac{\partial V_z(x)}{\partial x}-V_z(x)=dx\frac{\partial V_z(x)}{\partial x}
\end{equation*}
and
\begin{gather*}
[N_z(x+dx)-N_z(x)]=N_x(x+dx)\tan \Theta _{x+dx}-N_x(x)\tan \Theta _{x}\\
=N_x(x+dx)\frac{\partial w(x+dx)}{\partial x}-N_x(x)\frac{\partial w(x)}{\partial x}\\
=\left(N_x(x)+dx\frac{\partial N_x(x)}{\partial x}\right)\left(\frac{\partial w(x)}{\partial x}+dx\frac{\partial ^2 w(x)}{\partial x^2}\right)-N_x(x)\frac{\partial w(x)}{\partial x}\\
=N_x(x)dx\frac{\partial ^2 w(x)}{\partial x^2}
\end{gather*}
where we used the fact from above that $(\partial N_x(x))/(\partial x)=0$. Hence we get
\begin{equation}
\label{shear}
\frac{\partial V_z(x)}{\partial x}+N_x(x)\frac{\partial ^2 w(x)}{\partial x^2}\equiv 0
\end{equation}
Considering the moment along y-direction one ends up with
\begin{gather*}
[-M_y(x+dx)+M_y(x)]+\left[\frac{dx}{2}V_z(x+dx)-\frac{(-dx)}{2}V_z(x)\right]=0\\
-M_y(x)-dx\frac{\partial M_y(x)}{\partial x}+M_y(x)+\frac{dx}{2}V_z(x)+\frac{(dx)^2}{2}\frac{\partial V_z(x)}{\partial x}+\frac{dx}{2}V_z(x)=0
\end{gather*}
Since the second order term in $dx$ is negligibly small
\begin{equation}
\label{moment}
\frac{\partial M_y(x)}{\partial x}-V_z(x)\equiv 0
\end{equation}

\subsubsection{Equation of motion}
\label{seqm}
The equation of motion for flexural motion in z-direction is given by
\begin{equation}
\label{eom}
\rho A\,dx\frac{\partial ^2 w(x,t)}{\partial t^2}=\sum F_z
\end{equation}
where the $F_z$'s are the respective forces in z-direction. Including the findings from above yields
\begin{gather*}
\rho A\frac{\partial ^2 w(x,t)}{\partial t^2}=\frac{\partial V_z(x,t)}{\partial x}+N_x(x,t)\frac{\partial ^2 w(x,t)}{\partial x^2}\\
=\frac{\partial ^2 M_y(x,t)}{\partial x^2}+N_x(x,t)\frac{\partial ^2 w(x,t)}{\partial x^2}
\end{gather*}
Using \eq{mz} and \eq{newnx} the equation of motion can be expressed as
\begin{gather*}
\rho A\frac{\partial ^2 w(x,t)}{\partial t^2}=b\left\{\widetilde B\frac{\partial ^2}{\partial x^2}\left[\frac{\partial u(x,t)}{\partial x}+\frac{1}{2}\left(\frac{\partial w(x,t)}{\partial x}\right)^2\right]-\widetilde D\frac{\partial ^4 w(x,t)}{\partial x^4}+N_x\frac{\partial ^2 w(x,t)}{\partial x^2}\right\}
\end{gather*}
We already know that $\widetilde B=0$ due to choice of coordinate system. For the special case of a vibrating doubly clamped string, the following simplifying assumptions can additionally be made:
\begin{itemize}
\item no axial displacement of the cross-section \eq{ux}:\\
$\implies u(x)=0$
\item linearization of the strain \eq{epsilon}:\\
$\implies \frac{1}{2}((\partial w)/(\partial x))^2=0$
\end{itemize}
If we apply these assumptions to the equation of motion we get
\begin{equation}
\label{reseom}
\rho A\frac{\partial ^2 w(x,t)}{\partial t^2}=-b\left[\widetilde D\frac{\partial ^4 w(x,t)}{\partial x^4}-\widetilde N\frac{\partial ^2 w(x,t)}{\partial x^2}\right]
\end{equation}
This can be expressed as
\begin{equation}
\label{fineom}
\rho A\frac{\partial ^2 w(x,t)}{\partial t^2}=-D\frac{\partial ^4 w(x,t)}{\partial x^4}+ N\frac{\partial ^2 w(x,t)}{\partial x^2}
\end{equation}
using the effective flexural rigidity
\begin{equation}
\label{D}
D=b\widetilde D=b\frac{E_1^2d_1^4+2E_1E_2d_2(2d_1^3+2d_1d_2^2+3d_1^2d_2)+E_2^2d_2^4}{12(E_1d_1+E_2d_2)}=E_1I_{\rm eff}^{\rm oop}=(EI)_{\rm eff,oop}
\end{equation}
where $I_{\rm eff}^{\rm oop}$ is the effective area moment of inertia (see e.g. Gere et al.\cite{Mech_Mat2} for a rigorous derivation) and
\begin{equation}
\label{N}
N=b\widetilde N=b(\sigma _1d_1+\sigma _2d_2)=(\sigma A)_{\rm eff}
\end{equation}
the effective residual force, where $A=b(d_1+d_2)$ is the cross section area of the bilayer system. Defining the effective density as
\begin{equation}
\label{rho}
(\rho A)_{\rm eff}=\frac{\rho _1d_1+\rho _2d_2}{d_1+d_2}A=\widetilde{\rho}A
\end{equation}
one can rewrite the equation of motion as
\begin{equation}
\label{eb}
(\rho A)_{\rm eff}\frac{\partial ^2 w(x,t)}{\partial t^2}=-(EI)_{\rm eff,oop}\frac{\partial ^4 w(x,t)}{\partial x^4}+(\sigma A)_{\rm eff}\frac{\partial ^2 w(x,t)}{\partial x^2}
\end{equation}
This equation has the form of an Euler-Bernoulli equation incorporating tensile stress with effective parameters. For a doubly clamped resonator the corresponding boundary conditions are fixed-fixed, i.e.
\begin{equation*}
w(0)=0,\, w(L)=0,\, \frac{\partial w(0)}{\partial x}=0,\, \frac{\partial w(L)}{\partial x}=0
\end{equation*}
The solution in the presence of strong tensile residual stress can be approximated by using simply supported boundary conditions\cite{timoshenko2}, i.e.
\begin{equation*}
w(0)=0,\, w(L)=0,\, \frac{\partial ^2w(0)}{\partial x^2}=0,\, \frac{\partial ^2w(L)}{\partial x^2}=0
\end{equation*}
In this case the equation of motion is solved by
\begin{equation}
\label{sol}
w(x,t)=C\sin \left(\frac{j\pi x}{L}\right)\exp (i\omega _jt)
\end{equation}
where $j=1,2,...$ is the harmonic mode index with angular frequency
\begin{equation}
\label{w0}
\omega_{j}=\frac{j^2\pi ^2}{L^2}\sqrt{\frac{(EI)_{\rm eff}}{(\rho A)_{\rm eff}}}\sqrt{1+\frac{(\sigma A)_{\rm eff}L^2}{j^2(EI)_{\rm eff}\pi ^2}}
\end{equation}
corresponding to a frequency $f=\omega /(2\pi)$
\begin{equation}
\label{f}
f_{j}=\frac{j^2\pi}{2L^2}\sqrt{\frac{(EI)_{\rm eff}}{(\rho A)_{\rm eff}}}\sqrt{1+\frac{(\sigma A)_{\rm eff}L^2}{j^2(EI)_{\rm eff}\pi ^2}}
\end{equation}

\subsection{In-plane vibration}
The in-plane vibration of a bilayer system along the y-direction follows basically the same assumptions and derivation as the out-of-plane vibration along the z-direction substituting the displacement $w(x,t)$ by $v(x,t)$. Nevertheless several expressions during the calculations change their structure which will be discussed in this section.

\subsubsection{Neutral plane}
The neutral plane changes it's position compared to the out-of-plane case. Using a coordinate system in which the neutral axis along the z direction has the value derived for the out-of-plane motion and coincides with the x-axis (i.e. $y=0$ and $z=0$)
\begin{gather*}
-\int_{area\,A_1}E_1\frac{y}{\kappa}\,dA_1-\int_{area\,A_2}E_2\frac{y}{\kappa}\,dA_2=0\\
=-\frac{E_1}{\kappa}\int_{-b/2}^{b/2}y\,dy\int_{-\bar z}^{d_1-\bar z} dz -\frac{E_2}{\kappa}\int_{-b/2}^{b/2}y\,dy\int_{d_1-\bar z}^{d_1+d_2-\bar z}dz\\
=-\frac{1}{\kappa}\left\{E_1[z]_{-\bar z}^{d_1-\bar z}[\frac{1}{2}y^2]_{-b/2}^{b/2}+E_2[z]_{d_1-\bar z}^{d_1+d_2-\bar z}[\frac{1}{2}y^2]_{-b/2}^{b/2}\right\}\\
=-\frac{1}{\kappa}\left[(E_1d_1+E_2d_2)\left(\frac{1}{2}\frac{b^2}{4}-\frac{1}{2}\frac{b^2}{4}\right)\right]=0
\end{gather*}
Hence the designated axis of symmetry is indeed the neutral axis.

\subsubsection{Strain}

For simplicity we use the simplifying linearized definition derived in section \ref{seqm}. Therefore we get right away
\begin{gather}
\sigma_{1x}=-E_1y\frac{\partial ^2v(x)}{\partial x^2}+\sigma _1 \,\,\,\,\, {\rm for}\,\, -\bar z < z < d_1-\bar z\\
\sigma_{2x}=-E_2y\frac{\partial ^2v(x)}{\partial x^2}+\sigma _2 \,\,\,\,\, {\rm for}\,\, d_1-\bar z < z < d_1+d_2-\bar z
\end{gather}

\subsubsection{Forces and moments}

We derive the axial force analogously to the out-of-plane case

\begin{gather*}
N_x=\int_{area\,A_1}\sigma _{1x}\,dA+\int_{area\,A_2}\sigma _{2x}\,dA\\
=\int_{-b/2}^{b/2}\sigma _{1x}\,dy\int_{-\bar z}^{d_1-\bar z}\,dz+\int_{-b/2}^{b/2}\sigma _{2x}\,dy\int_{d_1-\bar z}^{d_1+d_2-\bar z}dz\\
=\int_{-b/2}^{b/2}\left[-E_1y\frac{\partial ^2v(x)}{\partial x^2}+\sigma _1\right]\,dy\,d_1+\int_{-b/2}^{b/2}\left[-E_2y\frac{\partial ^2v(x)}{\partial x^2}+\sigma _2\right]\,dy\,d_2\\
=-(E_1d_1+E_2d_2)\frac{\partial ^2v(x)}{\partial x^2}\left(\frac{1}{2}\frac{b^2}{4}-\frac{1}{2}\frac{b^2}{4}\right)+b(\sigma_1 d_1+\sigma _2d_2)\\
=b(\sigma_1 d_1+\sigma _2d_2)
\end{gather*}
Again defining the residual force $\widetilde N$
\begin{equation}
\label{nxin}
N_x=b(\sigma_1 d_1+\sigma _2d_2)=b\widetilde N=N=(\sigma A)_{\rm eff}
\end{equation}
The bending moment now in z-direction reads as previously
\begin{gather*}
M_z=\int_{area\,A_1}(\vec{r}\times \vec{\sigma})_z\,dA+\int_{area\,A_2}(\vec{r}\times \vec{\sigma})_z\,dA\\
=-\int_{-b/2}^{b/2}y\sigma _{1x}\,dy\int_{-\bar z}^{d_1-\bar z}dz-\int_{-b/2}^{b/2}y\sigma _{2x}\,dy\int_{d_1-\bar z}^{d_1+d_2-\bar z}dz\\
=-\int_{-b/2}^{b/2}y\left[-E_1y\frac{\partial ^2v(x)}{\partial x^2}+\sigma _1\right]\,dy\,d_1-\int_{-b/2}^{b/2}y\left[-E_2y\frac{\partial ^2v(x)}{\partial x^2}+\sigma _2\right]\,dy\,d_2\\
=(E_1d_1+E_2d_2)\frac{\partial ^2v(x)}{\partial x^2}\left(\frac{1}{3}\frac{b^3}{8}+\frac{1}{3}\frac{b^3}{8}\right)\\
=\frac{b^3}{12}(E_1d_1+E_2d_2)\frac{\partial ^2v(x)}{\partial x^2}
\end{gather*}
together with the tension stiffness $\widetilde A$
\begin{equation}
\label{ain}
M_z=\frac{b^3}{12}(E_1d_1+E_2d_2)\frac{\partial ^2v(x)}{\partial x^2}=\frac{b^3}{12}\widetilde A \frac{\partial ^2v(x)}{\partial x^2}
\end{equation}

\subsubsection{Equilibrium equations}
The forces along the x-direction remain the same as in the out-of-plane case. The same holds for forces along the y-direction, considering a shear force $V_y$ due to symmetry reasons, therefore
\begin{equation}
\label{inshear}
\frac{\partial V_y(x)}{\partial x}+N_x(x)\frac{\partial ^2 v(x)}{\partial x^2}\equiv 0
\end{equation}
yields an expression of the same structure as \eq{shear}.
For the bending moment one has to keep in mind that the sign of $M_z$ opposes that of $M_y$ for the out-of-plane mode (compare \eq{ain} and \eq{mz} with $\widetilde B=0$) originating from the vector product $(\vec{r}\times \vec{\sigma})_z$. Consequently we write the equilibrium equation for the bending moment along the z-direction
\begin{gather*}
[M_z(x+dx)-M_z(x)]+\left[\frac{dx}{2}V_y(x+dx)-\frac{(-dx)}{2}V_y(x)\right] \equiv 0
\end{gather*}
With the same arguments as for the out-of-plane mode one gets
\begin{equation}
\frac{\partial M_z(x)}{\partial x}+V_y(x)\equiv 0
\end{equation}

\subsubsection{Equations of motions}

Using the above findings allows to write down the equation of motion analogously to the out-of-plane case. We obtain
\begin{equation}
\label{ineom}
\rho A\frac{\partial ^2 v(x,t)}{\partial t^2}=-D\frac{\partial ^4 v(x,t)}{\partial x^4}+ N\frac{\partial ^2 v(x,t)}{\partial x^2}
\end{equation}
again an Euler-Bernoulli type differential equation with effective, yet modified elastic parameters:
\begin{equation}
(\rho A)_{\rm eff}\frac{\partial ^2 v(x,t)}{\partial t^2}=-(EI)_{\rm eff,ip}\frac{\partial ^4 v(x,t)}{\partial x^4}+(\sigma A)_{\rm eff}\frac{\partial ^2 v(x,t)}{\partial x^2}
\end{equation}
Like \eq{eb} it is solved by frequency solutions of the form
\begin{equation}
\label{fin}
f_{j}=\frac{j^2\pi}{2L^2}\sqrt{\frac{(EI)_{\rm eff}}{(\rho A)_{\rm eff}}}\sqrt{1+\frac{(\sigma A)_{\rm eff}L^2}{j^2(EI)_{\rm eff}\pi ^2}}
\end{equation}
but now with the effective elastic parameters for the in-plane oscillation, the effective flexural rigidity
\begin{equation}
\label{Din}
D=b\widetilde D=\frac{b^3}{12}(E_1d_1+E_2d_2)=E_1I_{\rm eff}^{\rm ip}=(EI)_{\rm eff,ip}
\end{equation}
where $I_{\rm eff}^{\rm ip}$ is the effective area moment of inertia. The effective residual force
\begin{equation}
\label{Nin}
N=b\widetilde N=b(\sigma _1d_1+\sigma _2d_2)=(\sigma A)_{\rm eff}
\end{equation}
has the same value as for the out-of-plane vibration.

\section{Higher harmonic modes}

We investigated the influence of metallization layers on silicon nitride string resonators not only for the fundamental flexural modes as described in the main article. In order to show that our theoretical model correctly describes the previous findings for non-metallized nanomechanical string resonators\cite{2006_Verbridge_JAP2, 2010_Quirin_PRL2}, we recorded a series of higher harmonic flexural modes of the bilayer system. Figure\,S4 depicts the inverse quality factor of the second harmonic ($j=2$) in-plane flexural mode as a function of the gold film thickness. Again as in Fig.\,3 of the main article we fitted a linear function to the first five data points in the approximation of small gold thickness.

\begin{figure}[htb]
\label{s4}
\includegraphics[scale=1]{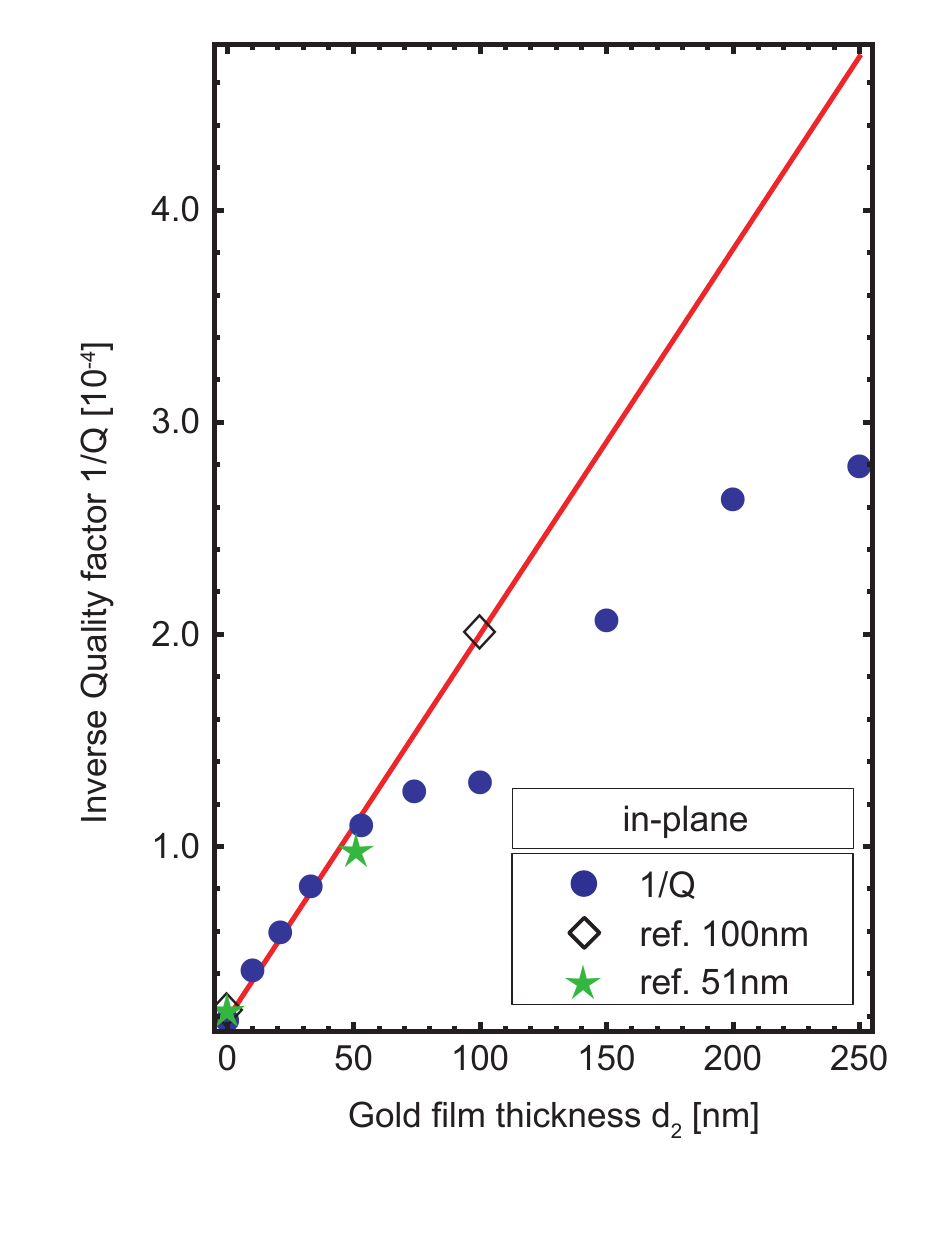}
\caption{Inverse quality factor 1/Q versus gold film thickness $d_2$ of the in-plane first harmonic resonance ($j=2$) of a 53\,\textmu m long resonator. Red solid line shows linear fit of the first 5 data points. Green stars and open diamonds correspond to reference samples with 51\,nm and 100\,nm gold film as described in the main text, respectively.}
\end{figure}

Here, it is immediately apparent that the linear approximation breaks down for higher gold thickness. The increasing deviation for larger $d_2$ is much more pronounced than for the fundamental modes described in the main text. We attribute this to the effect of frequency dependent, defect-mediated damping. Consistently the metal's quality factor $Q_m$ shows lower values (Fig.\,S5) than for the fundamental mode (compare main text), but also tends to qualitatively increase approximately linear with frequency (or inverse length)\cite{2006_Verbridge_JAP2}.
\begin{figure}[htb]
\label{5}
\includegraphics[scale=0.4]{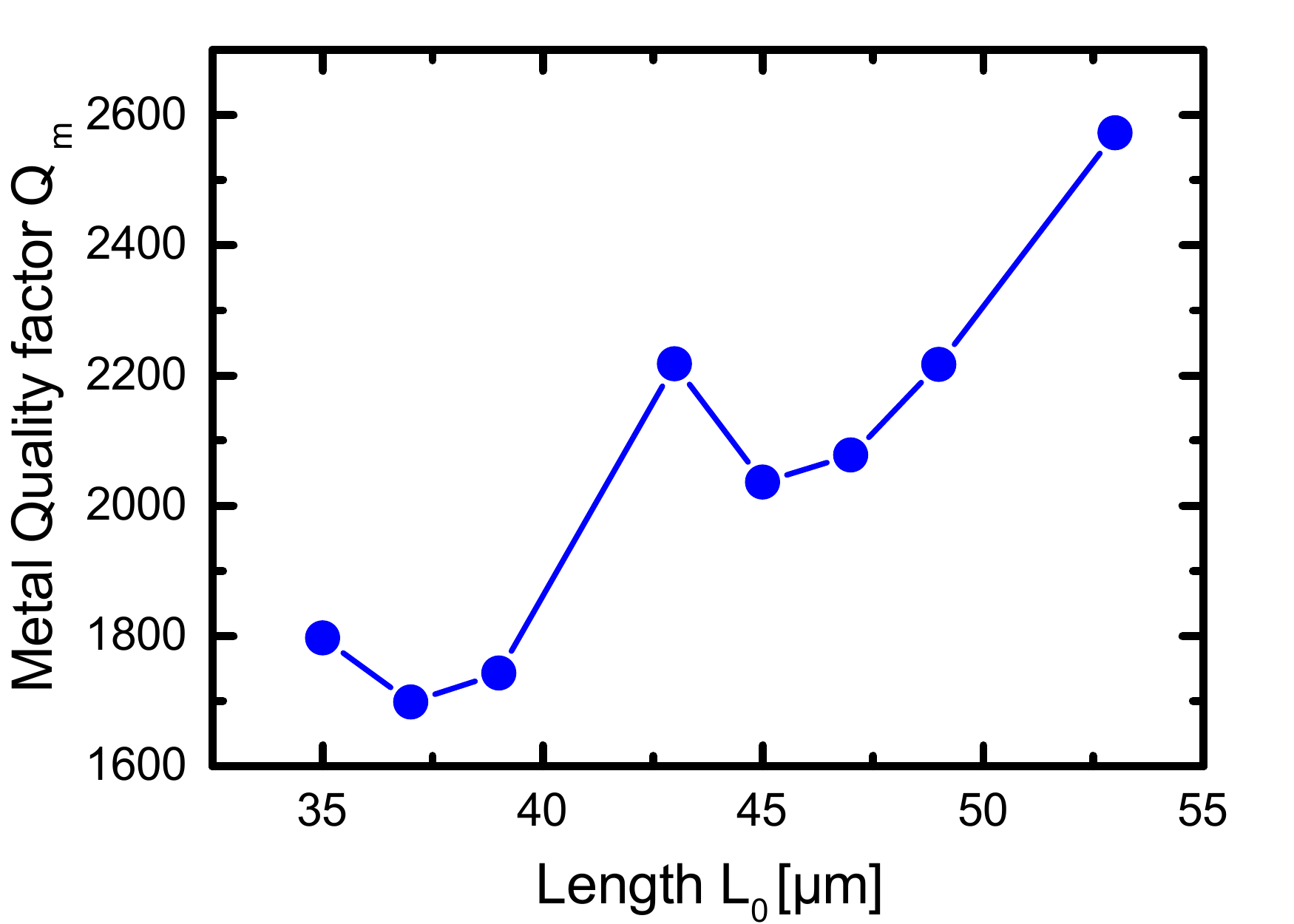}
\caption{Metal quality factor $Q_m$ of the in-plane second harmonic resonance ($j=2$) versus resonator length.}
\end{figure}
Note that resonators of arbitrary lengths may inhere different intrinsic quality factors due to fabrication imperfections of the respective resonator.
To further detail the frequency dependence of $Q_m$ we plot the obtained metal quality factor $Q_m$ of the flexural out-of-plane mode of a 53\,\textmu m long resonator for different harmonic modes (Fig.\,S6) along with the Q-factors of the bare, unmetallized SiN string. The observed linear decrease in $Q_m$ confirms not only that $Q_m$ is indeed limited by defect-mediated damping, but also suggests that, at least for this particular resonator, the evaporation produced a quite homogenous gold film. Otherwise the quality factor should change due to inhomogenous nucleation with respect to the mutual node and antinode position in the vibration of different harmonics\cite{2012_Chaste_NatNano2}.

\begin{figure}[htb]
\label{6}
\includegraphics[scale=0.4]{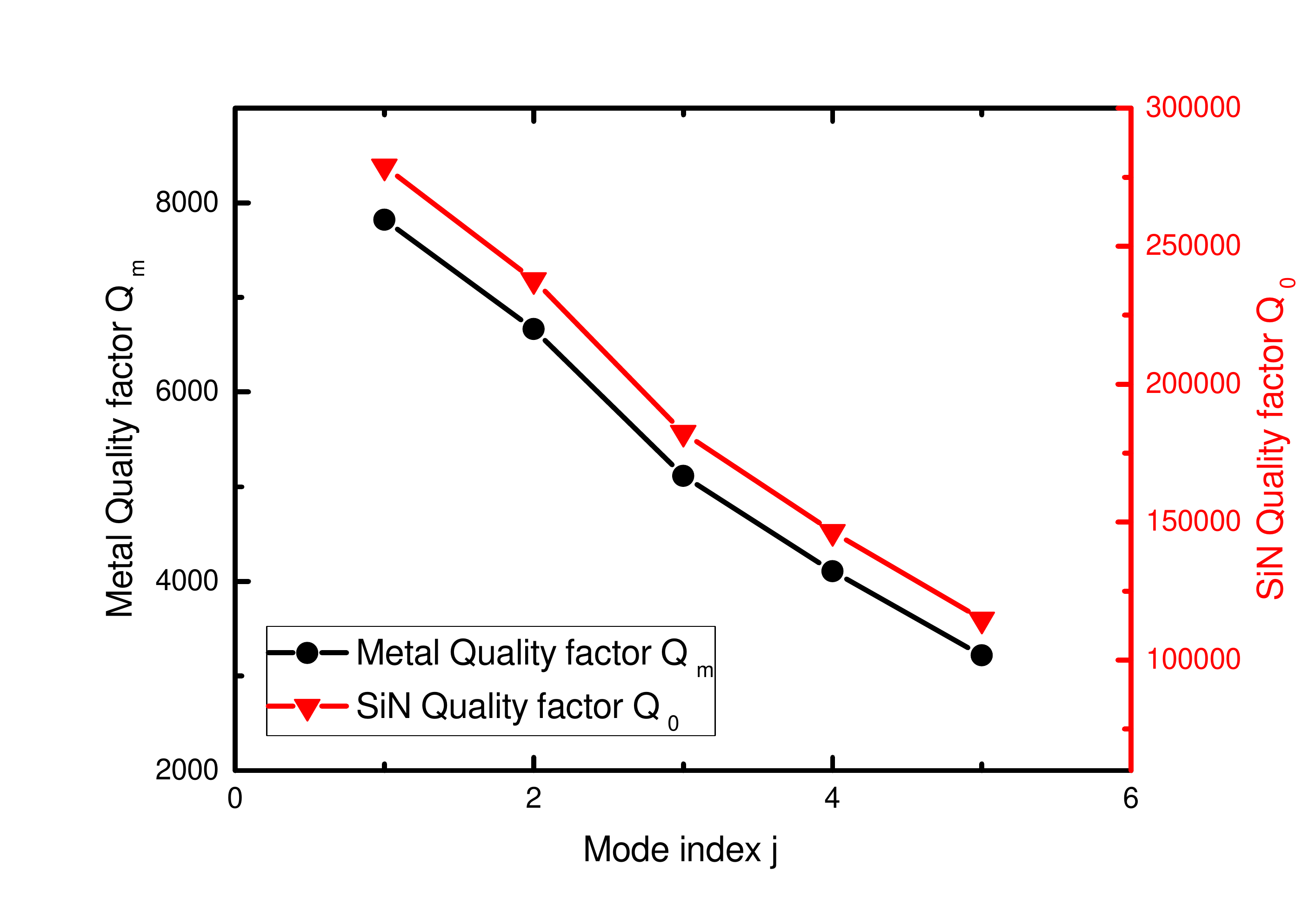}
\caption{Metal quality factor $Q_m$ versus harmonic mode index of the fundamental out-of-plane mode of a 53\,\textmu m long resonator (black dots, left scale). Red triangles depict non-metallized quality factors $Q_0$ of the silicon nitride string (right scale).}
\end{figure}

\section{Effective resonator length}

As described in the main text the resonator's clamping points exhibit a finite undercut resulting from the isotropic chemical wet etching step releasing the resonator. This undercut slightly changes the resonator's effective length. We find this effect to be dependent on the flexural mode polarization as well as the aspect ratio. Hence, fitting equation\,(1) from the main text to the experimental data (Fig.\,2) requires not only the tensile stress but also the resonator's length as an open fit parameter. While the obtained values for $\sigma _2$ are discussed in the inset of Fig.\,2 of the main text, the resulting values for $L$ are shown in Table\,S1 for the fundamental in- and out-of-plane flexural mode including the errors of the fits. For the case of longer resonators the effective length deviates only slightly from the nominal length, whereas this deviation gets more pronounced for decreasing length and therefore aspect ratio. We associate this to the fact that clamping losses, and therefore the influence of the undercut, become more dominant with decreasing aspect ratio\cite{2006_Verbridge_JAP2}.

\begin{table}[!htb]

\begin{center}
\begin{tabular}{|c|c|c|c|c|c|c|c|c|c|}
\hline
Nominal Length $L_0$ [\textmu m] & 53 & 49 & 47 & 45 & 43 & 39 & 37 & 35 & 33\\
\hline
Effective in-plane length [\textmu m] & 51.25 & 49.12 & 48.14 & 47.04 & 46.10 & 44.02 & 43.02 & 41.99 & not found\\
Fit Error [\textmu m] & 0.34 & 0.35 & 0.34 & 0.33 & 0.33 & 0.31 &  0.31 & 0.30 & not found\\
\hline
Effective out-of-plane length [\textmu m] & 53.90 & 51.74 & 50.76 & 49.67 & 48.70 & 46.63 & 45.66 & 44.64 & 43.76\\
Fit Error [\textmu m] & 0.37 & 0.39 & 0.38 & 0.36 & 0.36 & 0.35 & 0.35 & 0.34 & 0.34\\
\hline
\end{tabular}
\end{center}
\caption{\label{tabledue} Fit parameters for the effective resonator length obtained from the fit of equation\,(1) of the main text to the measured resonant frequencies vs. gold film thickness (Fig.\,2) for in- and out-of-plane fundamental modes.}
\end{table}

\end{document}